\documentstyle[a4,12pt,amssymb]{article}
\begin{document}

\def\reflist{\section*{References\markboth
        {REFLIST}{REFLIST}}\list
        {[\arabic{enumi}]\hfill}{\settowidth\labelwidth{[999]}
        \leftmargin\labelwidth
        \advance\leftmargin\labelsep\usecounter{enumi}}}
\let\endreflist\endlist \relax

\def\a{\alpha}
\def\b{\beta}
\def\g{\gamma}
\def\d{\delta}
\def\e{\epsilon}
\def\Tr{{\rm Tr \,}}
\newcommand{\be}{\begin{equation}}
\newcommand{\ee}{\end{equation}}
\newcommand{\ba}{\begin{eqnarray}}
\newcommand{\ea}{\end{eqnarray}}
\newcommand{\dal}{\raisebox{0.085cm}
{\fbox{\rule{0cm}{0.07cm}\,}}}
\newcommand{\dslash}{{\not\!\partial}}
\catcode`\@=11
\titlepage
\begin{flushright}
LPT-ORSAY 01-96 \\
CPTH-S044.1001 \\
LPTM 01-33
\end{flushright}
\vskip 1cm
\begin{center}
{\Huge \bf D-branes in String theory Melvin backgrounds}
\end{center}
\vskip 1cm
\begin{center}

 E. Dudas$^{\,a, b}$  and J. Mourad$^{\,a, c}$
\end{center}
\vskip 0.5cm
\begin{center}

$^a$ Laboratoire de Physique Th{\'e}orique
\footnote{Unit{\'e} Mixte de Recherche du CNRS (UMR 8627).}, \\
Universit{\'e} de Paris-Sud, B{\^a}t. 210, F-91405 Orsay Cedex\\
$^b$ Centre de Physique Th{\'e}orique, \\
Ecole Polytechnique, F-91128 Palaiseau\\
$^c$ Laboratoire de Physique Th{\'e}orique et Mod{\'e}lisation,\\
Universit{\'e} de Cergy-Pontoise, Site de Neuville III, F-95031
Cergy-Pontoise
\end{center}
\vskip 2cm

\begin{center}
{\large Abstract}

\end{center}

We determine the consistent D-brane configurations in type II
nonsupersymmetric Melvin Background. The D-branes are analysed from 
three complementary points of view: the effective Born-Infeld
action, the open string partition function and 
the boundary state approach. We show the agreement of the results
obtained by the three different approaches. Among the surprising features 
is the existence of supersymmetric branes, some of them 
having a quasi-periodic direction.
We also discuss the generalisation to  backgrounds with  several
magnetic fields, some of them preserving in the closed and the open
spectra some amount of supersymmetry. 
The case of rational magnetic flux, equivalent to freely-acting
noncompact orbifolds, is also studied. It allows more 
possibilities of consistent D-brane configurations.

\newpage
\section{Introduction and Summary of results}

D-branes \cite{Dai:1989ua,Polchinski:1995mt}
 are an essential ingredient in string theory.
In addition to their initial role in testing
conjectures on the non-perturbative regime of supersymmetric
string theories, they proved to be essential tools to
study the spacetime geometry at substringy distances 
\cite{Douglas:1997yp} and to holographically relate theories with gravity
to Yang-Mills theories on lower dimensional spacetimes 
\cite{Aharony:2000ti}. The brane world scenario in its new version
represents an important phenomenological developpement inspired by
D-branes.  All these issues   motivate 
the determination of the D-brane content
of a string theory in a general background.
Many authors have studied D-branes on compact Lie groups or 
their quotients using the exactly solvable WZWN models
\cite{Cardy:1989ir}-\cite{ Maldacena:2001ky,Dudas:2001wd}.
The D-branes in non-compact spacetime were considered for non-compact 
orbifolds or some group manifolds \cite{Bachas:2001bt}.
The aim of this paper
is the  study of D-branes on another interesting non-trivial
and exactly solvable background: the Melvin spacetime
\cite{Melvin} - \cite{Takayanagi:2001jj}.

The Melvin background is a flat space-time $R^{9} \times S^1$
subject to the twisted identification (I) $(y,Z_0)=(y+2\pi R, Z_0e^{i2\pi R
B})$. A constant spinor in $R^{9} \times S^1$ is not invariant
under (I), it picks a phase $e^{\pm i\pi RB }$ and so 
all the supersymmetries are broken. If we consider the
identification (II) $(y,Z_0,W_0)= (y+2\pi R, Z_0e^{i2\pi R
B_1}, W_0e^{i2\pi R B_2})$
then the phase may be compensated when $B_1=\pm B_2$
leaving half of the initial supersymmetries.
That the Melvin Background is obtained from flat spacetime
by an identification strongly suggests its solvability
which we review in Section II. In that section 
we also determine the quantization of the open strings and 
the various possibilities of consistent boundary conditions from the 
sigma model approach. One of the interesting
boundary conditions is Dirichlet for $Z_0$ and Neumann for $y$. 
In this particular case, the intersection of this curve with the 
$(y, Z_0)$ Melvin space, when $BR$ is not rational, is a non-compact 
line. The momenta along this 
direction are however quantized in terms of two integers, defining a
quasi-periodic space\footnote{An example of a quasi-periodic space
is an infinite length geodesic 
(which is not parallel to one of the cycles) on a  torus.}. The existence
of a mass gap in the spectrum in a noncompact space is a remarkable
phenomenon with possible phenomenological implications.
In section III,
the low energy couplings of the D-branes to the closed spectrum are
examined from the point of view of the Born-Infeld effective action.
In Section IV we determine the cylinder partition functions
from the one-loop open string amplitude for various types of D-branes,
parallel or orthogonal to the relevant coordinates of the Melvin geometry. 
This allows the determination of the D-brane spectra. Some of them have
open string Landau levels and some do not. Similarly, the boundary
conditions on D-branes fix the 
couplings of D-branes to the Landau levels of the closed states. 
This is determined
by the position of the D-brane with respect to the Melvin geometry.    
Depending on the case, supersymmetry can be completely broken, can be
present in the massless spectrum or can even be present in the complete 
(massive and massless) spectrum of D-branes. 
  
The results of our work may shed light on the fate of closed-string
tachyons in curved backgrounds, in the presence of D-branes. Indeed,
backreaction of D-branes studied in this paper on the curved (Melvin)
geometry could lift in mass the closed tachyon and stabilize
the background. It would be also interesting to study 
nonperturbative aspects of gauge theory on this type of D-branes
from the dual, curved gravitational background viewpoint.   
 
In Section V we verify the consistency of these amplitudes
by determining the boundary states of the D-branes
and checking the correct interpretation of the cylinder amplitude
as  a tree level closed string exchange.
We show that there are some cases which seem consistent from the
sigma model approach but which do not fulfill the requirements of
a correct dual open-closed channels interpretation.
In Section VI we study the case of the identification (II) 
which is supersymmetric. In Section VII we briefly examine the novel
features which arise when $BR$ is rational. The D-brane properties
in this case are different and can be understood as freely-acting 
interpolations
between flat-space D branes and D-branes in noncompact orbifolds.
Finally, an appendix contains the definitions and modular
transformation properties of the theta functions used in the text.
\section{Closed and open strings in the Melvin background}

The Melvin background is a flat space-time $R^9 \times S^1$ in which a
$2 \pi$ rotation in a compact coordinate $y$ is accompanied by a rotation
in a two-plane of polar coordinates $(\rho,\phi)$
\be
(y,\phi_0)=(y+2\pi R , \phi_0+2\pi RB ),\label{one}
\ee
where $\phi_0$ is an angle
\be
\phi_0=\phi_0+2\pi,\label{two}
\ee
and $B$ is the twist parameter, which can be seen as a closed string
magnetic field.
Due to (\ref{two}), $B R$ and $B R+n$ are equivalent
so we can suppose that $BR$ is between $0$ and $1$.
The metric is given by
\be
ds^2=d\rho^2+\rho^2d\phi_0^2+dy^2+d{\bf x}^2 \ ,
\ee
where ${\bf x}$ denotes 7 space-time coordinates.
If we define $\phi$ by $\phi=\phi_0-B  y$,
then it represents a true angular coordinate
and the metric is given by
\be
ds^2=d\rho^2+\rho^2(d\phi+B dy)^2+dy^2+d{\bf x}^2 \ .
\ee
Consider a sigma model with the Melvin background as a
target space. The world-sheet action reads
\be
S=- {{1}\over{4\pi\alpha'}}\int [ d\rho\wedge ^{*}d\rho+
\rho^2(d\phi+B dy)\wedge ^{*}(d\phi+B dy)
+dy\wedge ^*dy ] \ .
\ee
The equations of motion deriving from it read
\be
\partial_+\partial_-Z_0=0,\
\partial_+\partial_-y=0 \ , 
\ee
where we used the usual definitions $\sigma_{\pm} = \tau \pm \sigma$ and we
defined the free-field 
\be
Z_0=\rho \ e^{i(\phi+B y)}= Z \ e^{iB y} \ .
\ee
The solution to the field equations may be written as
\be
Z_0=Z_{0+}(\sigma_+)+Z_{0-}(\sigma_-), \
y=y_+(\sigma_+)+y_-(\sigma_-) \ , 
\ee
in terms of right and left moving free fields.
In order to get the explicit solution, one has to specify the boundary
conditions. The angular nature of $y$ and $\phi$
imposes that after quantization, their associated momenta are
quantized,
\be
P_y={k\over R} \ , \ P_{\phi}={L} \ , \label{quan}
\ee
where $k$ and $L$ are integers and
\be
P_\phi={{1}\over{2\pi\alpha'}}\int_0^{2\pi} d\sigma
\rho^2\left({d\phi \over d\tau}+B
{d y\over d\tau}\right), \
P_y={{1}\over{2\pi\alpha'}}
\int_0^{2 \pi} d\sigma {d y\over d\tau}\ +BP_\phi \ . \label{mom}
\ee
\subsection{ Closed strings}
For closed strings  the $Z$ coordinate is periodic and the
$y$ coordinate can have winding modes $n$ :
\be
Z(\sigma+2\pi,\tau)=Z(\sigma,\tau) \ , \  y(\sigma+2\pi)=y(\sigma)
+2\pi nR \ .
\ee
This implies for the free coordinate $Z_0$ the nontrivial boundary condition:
\be
Z_0(\sigma+2\pi)=e^{2\pi i B nR} Z_0(\sigma) \ .
\ee
The mode expansion  of the $y$ coordinate, due to (\ref{quan})
and (\ref{mom}),  is given by
\ba
y=y_0 + nR\sigma+\alpha'({k \over R}-B L)\tau+\sum_{m=1}^{\infty}
{1\over \sqrt{m}}&\Big[& {y_m }e^{-im\sigma_+}+
{y_m^\dagger}e^{im\sigma_+}\nonumber\\
&+&{\tilde y_m }e^{-im\sigma_-}+
{\tilde y_m^\dagger }e^{im\sigma_+}\Big].
\ea
The $y$ contribution to the   hamiltonian is
\be
N_y+\tilde N_y+{1 \over 2\alpha'}
[(nR)^2+(\alpha'({k \over R}-B L))^2]-{2
\over 24}.
\ee
The hamiltonian describing the $Z$ coordinate
in the winding sector $n$ is  the one of a twisted boson.
The mode expansion of $Z_0$ is thus given by
\ba
Z_0(\sigma,\tau)&=&\sqrt{\alpha'}\Big[
\sum_{m=1}^{\infty}{{a_m \over
\sqrt{m-\nu}}}e^{-i(m-\nu)\sigma_+}+\sum_{m=0}^{\infty}
{b_m^{\dagger} \over{\sqrt{m+\nu}}}e^{i(m
+\nu)\sigma_+} \nonumber\\
&+&  \sum_{m=0}^{\infty}{{\tilde a_m \over
\sqrt{m+\nu}}}e^{-i(m+\nu)\sigma_-}+\sum_{m=1}^{\infty}
{\tilde b_m^{\dagger} \over{\sqrt{m-\nu}}}e^{i(m
-\nu)\sigma_-}\Big],\label{expa}
\ea
where $\nu=f(B nR)$, the function $f$ is  1-periodic
and $f(x)=x$ for $0<x<1$. Explicitly $\nu=nBR-[nBR]$ when $n$ is
positive and  $\nu=nBR-[nBR]+1$ when $n$ is negative.
In (\ref{expa}), $a_n^\dagger$ and $b_n^\dagger$ are creation
operators.

The Hamiltonian
of the twisted boson, including the normal ordering constant, reads
\ba
H_\nu&=&{2 \over 24}-{2 \over 8}(1-2\nu)^2+
\nu \ (b_0^{\dagger}b_0+\tilde a^{\dagger}_0\tilde a_0)
\nonumber\\
&+&\sum_{m=1}^{\infty}(m-\nu)(a^{\dagger}_ma_m+
\tilde b^{\dagger}_m\tilde b_m)+ \sum_{m=1}^{\infty} 
(m+\nu)(\tilde a^{\dagger}_m\tilde a_m+ b^{\dagger}_m b_m) \ .
\ea
The angular momentum corresponding to a rotation
of the  $\phi$ coordinate is given by
\be
P_\phi\equiv J+\tilde J=\tilde a^{\dagger}_0\tilde a_0- b_0^{\dagger} b_0
+\sum_{m=1}^{\infty}( \tilde a_m^\dagger\tilde a_m
+ a_m^\dagger a_m-
\tilde b_m^\dagger\tilde b_m-b_m^\dagger b_m) \ . \label{angm}
\ee
The total Hamiltonian of the $y$ and $Z$ coordinates is thus  given by
\ba
H&=&L_0+{\bar L}_0 \nonumber\\
&=&N+\tilde N-\nu(J-\tilde J)+
{1 \over 2\alpha'}[(nR)^2+(\alpha'({k \over R}-B (J+\tilde J))^2]
-{1 \over 4}(1-2\nu)^2 \ , \label{mass}
\ea
where $N$ ($\tilde N$) denotes the total number of left (right) oscillators.
The momentum, corresponding to a translation in the world sheet
$\sigma$ coordinate, is
\be
P=L_0-\tilde L_0=N-\tilde N-nk+(nBR-\nu)(J+\tilde J) \ .
\ee
The formulae given above are valid when $\nu$ is not zero.
When $\nu$ vanishes, $Z$ acquires zero modes and the hamiltonian of the
$Z$ coordinates becomes
\be
H_0=-{4 \over 24} +{\alpha' \over 2}(p_1^2+p_2^2)+N_Z+\tilde N_Z \ .
\ee
In this case, the rotation generator acquires an orbital part:
\be
J+\tilde J=xp_y-yp_x+\sum_{m=1}^{\infty}( \tilde a_m^\dagger\tilde a_m
+ a_m^\dagger a_m-\tilde b_m^\dagger\tilde b_m-b_m^\dagger b_m).
\ee

The zero modes quantum numbers when $n \neq 0$ are given by
$(k,n,n_0,\tilde n_0)$, where $n_0, {\tilde n}_0$ will be interpreted as
Landau levels due to the presence of the magnetic field $B$.
The level matching condition
gives $-nk=[B Rn](\tilde n_0-n_0)$. This state has a
hamiltonian $(\alpha' /2) (k/R-B(\tilde n_0-n_0))^2+(nR)^2/(2\alpha')
+\nu(n_0+\tilde n_0)-(1-2\nu)^2/4.$

On the other hand, when $n=0$ the quantum numbers from the $Z$ coordinate
include the angular momentum $L$ and the momentum squared
($p^2=p_1^2+p_2^2$). The level matching condition is automatically
satisfied and the hamiltonian of the state is given by
$-6/24 +(\alpha' /2) ({k \over R}-B L)^2+\alpha' p^2 /2$.

The torus partition function, after a Poisson resummation on $k$,
is given by\footnote{The complete,modular invariant torus amplitude
is actually given, with this definition, by $\int 
(d^2 \tau / \tau_2) T$.}
\be
T={R\over \sqrt{\pi\alpha'\tau_2}}{1 \over |\eta|^2}
\sum_{\tilde k, n}e^{-{\pi R^2 \over \alpha'\tau_2}
|\tilde k+\tau n|^2}Tr_\nu\Big[e^{2\pi iRB \tilde k(J+\tilde J)}
q^{L_0}\bar q^{{\bar L}_0}\Big] \ ,
\ee
where $Tr_{\nu}$ is the trace in the sector twisted by $\nu$.
In order to calculate explicitly the traces in the above equation
let us note that when $\nu=0$ and $\tilde k=0$ the
trace over the zero modes involves an integration
on the noncompact two-dimensional momenta. On the other hand, when 
$\tilde k\neq 0$ the trace over the zero modes reads
\be
\int d^2 p \ <p|e^{ 2\pi i L B \tilde k R}|p> (q {\bar q})^{\alpha'
p^2 /4} = {1 \over \det
(1-\theta)}= {1 \over 4 (\sin \pi B R \tilde k)^2} \ , \label{zerom1}
\ee
where $\theta$ is a two-dimensional rotation by an angle
$2\pi B R \tilde k$.
Notice that the analogous contribution for the  case
of a compact two-torus replacing the $(\rho,\phi)$ plane, is one.
When $\nu=0$, the oscillators contribution is given by
\be
Tr_{\rm osc} [e ^{2\pi i{\tilde k} RB J}q^{L_0}]= 2\sin(\pi BR\tilde k)
{\eta \over \vartheta[^{1/2}_{1/2+RB\tilde k}]} \ . \label{tracezero}
\ee

The contribution of the $(y,\rho,\phi)$ coordinates to the
torus amplitude is thus explicitly given by
\be
T={R\over \sqrt{\pi\alpha'\tau_2}}{1 \over |\eta|^2}
\Big\{\sum_{\tilde k, n}e^{-{\pi R^2 \over \alpha'\tau_2}
|\tilde k+\tau n|^2}Z(\tilde k,w)\Big\} \ , \label{tor}
\ee
where (we suppose for the time being that $B R$ is not rational)
\be
Z(0,0)={V_2 \over 4\pi^2\alpha'\tau_2}{1 \over |\eta|^4} \ , \\
Z(\tilde k,n)=
\Big|{\eta \over \vartheta[^{1/2+RB n}_{1/2+RB \tilde k}]}
\Big|^2 \ , \ \forall (\tilde k, n) \neq (0,0) \  \label{part}
\ee
and in (\ref{part}) $V_2$ is the volume of the plane $(\rho,\phi)$.
There is a different and sometimes more illuminating
 way to obtain the partition function.
Consider the free theory on $R^3$ and then perform the orbifold
$R^3/Z$, where the action of the group
is generated by (\ref{one}). The untwisted part of the torus
amplitude is obtained by the insertion in the flat space amplitude
of the projector
onto states invariant under (\ref{one}). The latter  is given by
\be
\pi=\sum_{\tilde k} e^{-2\pi iR\tilde k P_y}e^{ 2\pi iRB \tilde k
P_\phi}. \label{proj}
\ee
The completion of the amplitude by modular invariance gives the
rest of the torus partition function.

The inclusion of the world-sheet fermions is straightforward.
The world-sheet superpartners of $Z_0$, (called $\lambda$ in what follows),
in the Ramond sector are twisted by $1-\nu$
and in the NS sector are twisted by $1/2-\nu$.
The total right and left moving hamiltonians are given by
\ba
L_0&=&N+{\alpha' \over 4}
\Big({k \over R}-B(J+\tilde J)+{nR\over \alpha'}\Big)^2
-\nu J +a \ , \nonumber\\
\bar L_0&=& {\tilde N} + {\alpha' \over 4}
\Big({k \over R}-B(J+\tilde J)-{nR\over \alpha'}\Big)^2
+\nu \tilde J +\tilde a \ ,
\ea
where $N$ and ${\tilde N}$ include now the fermionic oscillators and  
\be
a(NS)={\nu-1 \over 2} \ , \ a(R)=0 \ . 
\ee
The angular momentum $J$ has in addition to the bosonic
contribution (\ref{angm}) the fermionic one
\be
J_f+\tilde J_f=- {1 \over 2\pi}\int (\lambda^\dagger\lambda
+\tilde\lambda^\dagger\tilde \lambda) \ .
\ee
The torus amplitude for type IIB (IIA) superstring
in the Melvin background, encoding the GSO projections,
is given by
\be
T=V_7{R\over \sqrt{\pi\alpha'\tau_2}}{1 \over (4\pi^2
\alpha'\tau_2)^{7/2} |\eta|^{12}}
\Big\{\sum_{\tilde k, n}e^{-{\pi R^2 \over \alpha'\tau_2}
|\tilde k+\tau n|^2}Z(\tilde k,n)\Big\} \ , \label{tor2}
\ee
with
\ba
Z(\tilde k,n)&=&{1 \over 4} \Big\{\sum_{\alpha,\beta}\eta_{\alpha
\beta}e^{-2\pi i \beta BRn}{\vartheta^3[^\alpha_\beta] \over \eta^3}
{\vartheta[^{\alpha+RBn}_{\beta+RB\tilde k}]
\over \vartheta[^{1/2+RBn}_{1/2+RB\tilde k}]}\Big\}
\Big\{\sum_{\alpha,\beta}\bar \eta_{\alpha
\beta}e^{2\pi i\beta BRn}
{\bar \vartheta^3[^\alpha_\beta] \over \bar \eta^3}
{\bar \vartheta[^{\alpha+RBn}_{\beta+RB\tilde k}]
\over \bar\vartheta[^{1/2+RBn}_{1/2+RB\tilde k}]}\Big\} \ ,
\nonumber \\
Z(0,0)&=&{V_2 \over 16 \pi^2\alpha'\tau_2|\eta|^4}
\Big\{\sum_{\alpha,\beta}\eta_{\alpha
\beta}{\vartheta^4[^\alpha_\beta] \over \eta^4}
\Big\}
\Big\{\sum_{\alpha,\beta}\bar \eta_{\alpha
\beta}{\bar \vartheta^4[^\alpha_\beta] \over \bar \eta^4}
\Big\} \ ,
\label{parti}
\ea
where
$\eta_{\alpha\beta}=(-1)^{2\alpha+2\beta+4\alpha\beta}$
and $\bar \eta_{\alpha \beta}=\eta_{\alpha \beta}$
for IIB and $\bar \eta_{1/2\  1/2}=-\eta_{1/2\  1/2}$ for IIA.
Using a Jacobi identity it is possible to cast
(\ref{parti}) in the form
\be
Z(\tilde k,n)= \ \Big|{\vartheta_1^4(BR(\tilde k+n\tau)/2|\tau)
\over \eta^3 \vartheta_1(BR(\tilde k+n\tau)|\tau)}\Big|^2 \ ,
\ee
which is the form found when the Green-Schwarz form of the
world-sheet action is used \cite{Russo:1996ik}.

\subsection{ Open strings }

For open strings, the consistent boundary conditions
correspond to Neumann or Dirichlet for $y$ and $Z_0$.
They are obtained by the requirement that the boundary term 
arising from the variation of  the
sigma model action vanishes. The boundary term reads
\ba
&&\partial_\sigma y\delta y+\partial_\sigma X_0\delta X_0
+\partial_\sigma Y_0\delta Y_0=\partial_\sigma y\delta y+
\partial_\sigma \rho\delta \rho+\rho^2\partial_\sigma\phi_0\delta
\phi_0 \nonumber \\
&&=\partial_\sigma \rho\delta \rho+\rho^2
(\partial_\sigma\phi+B\partial_\sigma y)\delta\phi
+(\partial_\sigma y+ B \rho^2
(\partial_\sigma\phi+B\partial_\sigma y))\delta y.\label{bou}
\ea
The coordinate of physical relevance is $Z$
but, as is manifest from (\ref{bou}),
 this is not the coordinate with simple boundary conditions.
The sigma model action allows uncorellated boundary conditions
for $Z_0$ and $y$. In the following we examine the different
possibilities.

\begin{itemize}

\item[{\bf a}-] Neumann conditions for all the coordinates.
We have $Z_{0+}(\sigma)=Z_{0-}(\sigma)=Z_{0+}(\sigma+2\pi)$
and $y_+=y_-$, $y= \alpha'(k/R-B J)\tau+osc.$
So $Z_0$ represents a usual free untwisted world-sheet boson.
In particular, it has zero modes corresponding  to two noncompact
momenta and
\be
J=xp_y-yp_x+ \sum_{m=1}^{\infty}(
a_m^\dagger a_m- b_m^\dagger b_m) \ .
\ee
The difference with respect to free bosons is that the
Kaluza-Klein contribution to the Hamiltonian is replaced by
$(k/R-BJ)^2$.\\

\item[{\bf b}-] Dirichlet conditions for $y$ and 
$\phi_0$, and Neumann for
$\rho$. This implies that $y$ and $\phi$ are Dirichlet.
The expansion of $y$ reads
\be
y= y_0 + 2 \sigma nR +osc.,
\ee
where $n$ is the winding mode.
If $\phi(0)=\phi(\pi)=0$, then
\be
\phi_0(\pi)-\phi_0(0)=2B\pi nR \ .
\ee
This implies that $Z_{0+}$ and $Z_{0-}$ are related by
\be
Z_{0+}(\sigma)=Z_{0-}(\sigma)^* \ ,
\ee
and
\be
Z_{0+}(\sigma+2\pi)=Z_{0+}(\sigma)e^{4 \pi iB nR}.
\ee
The free field $Z_0$ is thus twisted by $\nu = 2BRn-[2BRn]$ and its
mode expansion reads
\be
Z_{0+}(\sigma_+)=\sqrt{\alpha'}\Big[
\sum_{m=1}^{\infty}{{a_m \over
\sqrt{m-\nu}}}e^{-i(m-\nu)\sigma_+}+\sum_{m=0}^{\infty}
{b_m^{\dagger} \over{\sqrt{m+\nu}}}e^{i(m +\nu)\sigma_+}\Big]
\ee
and the Hamiltonian for the $(\rho,\phi)$ coordinates is given by
\be
H_\nu={1 \over 24}-{1 \over 8}(1-2\nu)^2+
\nu b_0^{\dagger}b_0
+\sum_{m=1}^{\infty}(m-\nu)a^{\dagger}_ma_m
+\sum_{m=1}^{\infty} (m+\nu)
b^{\dagger}_m b_m \ .
\ee
Notice that with these boundary conditions, the quantization
condition on $P_y$ and $P_\phi$
as defined in (\ref{mom}) is no longer valid.
The full Hamiltonian is given by
$H=H_\nu+N_y+{1 \over \alpha'}(nR)^2-{1 \over 24}$.\\

\item[{\bf c}-] Dirichlet boundary condition for $y$, $\phi_0$ and
$\rho$. This corresponds to  Dirichlet boundary conditions for
$Z$. If $y$ has a winding mode $n$ and take for simplicity $y_0=0$, then
$Z_0(0,\tau)=Z(0,\tau)$ but $Z_0(\pi,\tau)=e^{iB2\pi nR}
Z(\pi,\tau)$. In particular if $Z(0,\tau)=Z(\pi,\tau)$
then $\phi_0(0)-\phi_0(\pi)= - 2\pi BnR$ and $|Z_0(0)-Z_0(\pi)|^2=
4 \rho_0^2 \sin^2(\pi BnR)$. This gives a contribution 
proportional to $\sin^2(\pi BRn)$
to be added to the free Hamiltonian.

\item[{\bf d}-] $y$ Neumann and $Z_0$ Dirichlet
with $Z_{0}(0,\tau)$ and $Z_0(\pi,\tau)$
different from zero. The identification (\ref{one})
leads to nontrivial zero modes for the $y$ coordinate.
In fact the subspace $\phi_0=const.$
is non-compact when $BR$ is irrational.
In order to determine the allowed $y$ momenta,
consider a function on the $y,\phi_0,\rho$ space.
It has the expansion
\be
f(y, \phi_0,\rho)=\sum_{k,l}e^{iy{k \over R}}
e^{il \phi} c_{kl}(\rho) \ ,
\ee
where we used the fact that the variables $y,\phi$ are periodic.
Now if we restrict this function to the subspace $\phi_0=c$,
we get the quasi-periodic function in $y$ 
\be
g(y)=  \sum_{k,l} e^{iy{k \over R}} e^{il (c-By)} c_{kl}(\rho) \ ,
\ee
that is the allowed momenta in the $y$ direction are 
characterised by two integers $k$ and $l$ and 
\be
p_y={k \over R}-lB \ . \label{quasi}
\ee
This is a novel kind of "compactification" on quasi-periodic
spaces. The length of the $y$ coordinate is not finite
but its momenta are quantized with two integers. The existence
of non-compact dimensions with however a mass gap in the spectrum
(\ref{quasi}) can have interesting phenomenological implications.

\item[{\bf e}-]  $y$ Neumann, $Z_0$ Dirichlet with $Z_0(0,\tau)=
Z_0(\pi,\tau)=0$. 
At the origin of the $(\rho, \phi_0)$ plane, the identification
(\ref{one}) does not lead to any twist. Indeed, the identification
(\ref{one}) $(y,Z_0)=(y+2\pi,Z_0e ^{2i\pi BR})$ shows clearly
the absence of any twist at the origin.
The hamiltonian in this case is nearly the same as the case a) above.
The difference is that $J$ does not have a zero mode component, that
is
\be
J= \sum_{m=1}^{\infty}(
a_m^\dagger a_m- b_m^\dagger b_m) \ .
\ee
This reflects in particular in the absence of the term (\ref{zerom1})
in the open string amplitude.

\item[{\bf f}-] $y$ Dirichlet and $Z_0$ Neumann.
There are winding modes for $y$. The hamiltonian of the system
is completely free.

\end{itemize}
\section{D-Branes in the Melvin background: Born-Infeld results}

A classical approach for determining properties of D-branes
uses the interaction of closed string fields with the
branes via the Born-Infeld action. In order to have a neat
field-theory interpretation for branes with Neumann boundary conditions
in $y$ in the Melvin background, we first
perform a Buscher T-duality in the
coordinate $y$. The resulting world-sheet action,
\be
S=- {{1}\over{4\pi\alpha'}}\int [ d\rho\wedge ^{*}d\rho+
{1 \over 1+ B^2 \rho^2} (\rho^2 d\phi \wedge ^{*} d\phi
+ d{\tilde y}\wedge ^*d{\tilde y} - 
2 B \rho^2  d \phi \wedge d {\tilde y}) ] \ , \label{bi1}
\ee
where ${\tilde y}$ denotes the T-dual coordinate, exhibits a curved
background as well as the presence of an antisymmetric tensor
$B_{\phi {\tilde y}}$. We determine in the following the
spectrum of classical fluctuations around this background and compare
with the closed string mass spectrum. By using these results in
the Born-Infeld action, one finds the one-point functions of closed
string states in the front of the D-branes, to be compared later on
with the result extracted from the string brane-brane amplitudes.
  The classical field equation for the dilaton $\Phi$, for example,
in the background (\ref{bi1}) reads
\be
{1 \over \rho} {\partial \over \partial \rho} (\rho
{\partial \Phi \over \partial \rho} ) + (1+B^2 \rho^2)
({\partial^2 \Phi \over \partial {\tilde y}^2} +
{1 \over \rho^2} {\partial^2 \Phi \over \partial \phi^2})+ \Delta_7
\Phi = 0 \ , \label{bi2}
\ee
where $\Delta_7$ denotes the d'Alembertian in the remaining (flat)
seven spacetime coordinates ${\bf x}$, which defines the mass of the 
particle $\Delta_7 \Phi = M^2 \Phi$.
Expanding the solution into Kaluza-Klein modes along ${\tilde y}$ and
angular momentum modes
\be
\Phi (\rho, \phi, {\tilde y},{\bf x}) = e^{i {k \over R} {\tilde y}+ i
  l \phi} \Phi_{k,l} (\rho, {\bf x}) \ , \label{bi3}
\ee
we find the Schr\"odinger-type equation
\be
- {1 \over \rho} {\partial \over \partial \rho} (\rho
{\partial \Phi \over \partial \rho} ) + ({l^2 \over \rho^2} + {k^2 B^2
\over R^2} \rho^2)
\Phi = (M^2 - {k^2 \over R^2} - B^2 l^2) \Phi \ . \label{bi4}
\ee
The equation (\ref{bi4}) is identical to the Schrodinger equation for
a two-dimensional oscillator with frequency $\omega_k = |k| B /
R$. The solution is easily found by making the change of variables
\be
\Phi = f(\omega_k \rho^2) \rho^{|l|} 
e^{- {\omega_k \rho^2 \over 2}} \ . \label{bi04}
 \ee
In terms of the function $f(z = \omega_k \rho^2)$, (\ref{bi4}) becomes
\be
z f^{''} + (1+|l|-z) f^{'} = -{1 \over 4 \omega_k} 
\left(M^2 - {k^2 \over R^2} - B^2 l^2 - 2 \omega_k (|l|+1)\right) f \ . \label{bi05}
\ee
The differential equation has normalizable solutions if and only if
\be
M^2 = {k^2 \over R^2} + B^2 l^2 + 2 \omega_k (2 j + |l| + 1) \ ,
\label{bi5}
\ee
for non-negative integer $j$. 
The mass operator (\ref{bi5}) indeed
coincides with the field-theory part ot the string mass operator
(\ref{mass}) with the following identifications
\be
l = n_0-{\tilde n}_0 \ , \ 2j+ |l|= n_0+ {\tilde n}_0+1 \ . \label{bi6}
\ee
The solution of (\ref{bi05}) gives $f(z) = A L_j^{|l|}(z)$, where $A$ is
a normalization constant and the $L$'s
are the Laguerre polynomials  
\be
L_j^{n} (z)=
{z^{-n} \over j !} e^{z} {d^j \over d z^j} (z^{j+n}
e^{-z}) \ . \label{bi7}
\ee
The normalisation of the wavefunction is obtained from 
the closed string effective action from the requirement of a
standard propagator.
The normalized (in flat space) dilaton wave-functions can be then written in the form
\be
\Phi_{k,l,j}(\rho,\phi,{\tilde y}) =  e^{i {k \over R} {\tilde y}+ i
 l \phi} \sqrt{\omega_k^{|l|+1} j ! \over \pi \ 
(j+|l|)!} \ \rho^{|l|} L_j^{|l|} (\omega_k \rho^2) 
\ e^{- {\omega_k \rho^2 \over 2}} \ . \label{bi8}
\ee

{\bf NN boundary conditions for $Z$}\\

If the D-brane under consideration has Neumann boundary conditions for
all relevant coordinates $y,Z$, after performing the T-duality on
$y$ described above let us put the brane at the point $y=0$ .
Then the Born-Infeld action reads
\be
S_{BI} = - T_p \int_{y=0} d^2 Z e^{- \Phi} \sqrt{ \det (g + B)} =
- T_p \int_{y=0} \rho d \rho d \phi \ e^{- \Phi} \ , \label{bi9}
\ee
where in deriving the last equality we used explicitly the
background (\ref{bi1}). The result is the same as if the
space were perfectly flat.
In order to compare field theory
considerations with string computations it is useful to extract the
one-point function $g_{k,l,j}$ of the dilaton in
front of the boundary (D-brane). By using (\ref{bi3}),(\ref{bi8}) 
we find
\be
g_{k,l,j}= 2 \delta_{l,0} \pi T_p \int \rho d \rho \ 
\Phi_{k,0,j} \ , \label{bi10}
\ee
and in particular the total angular momentum $l = n_0-{\tilde n}_0$ of closed
states coupling to the brane is zero. The explicit computation
of (\ref{bi10}) finally gives, by performing $j$ integration by parts,
the simple result 
\be
g_{k,0,j} = T_p \sqrt{1 \over \pi \omega_k} \int_0^{\infty}
dz \ L_j^{0}(z) \ e^{-{z \over 2}} =
2 (-1)^j \ T_p \sqrt{ \pi \over \omega_k} \ . \label{bi11}
\ee

{\bf DD boundary conditions for $Z$}\\

 The D-brane under consideration has Dirichlet boundary conditions for
the magnetized plane $Z$ and the D-brane is forced to sit at the origin
of the plane $Z=0$. After performing the T-duality on
$y$ and putting the brane at $y=0$, the Born-Infeld action reads
\be
S_{BI} = - T_p \int_{y=0} \delta^2 (Z) e^{- \Phi} \sqrt{ \det (g + B)} =
- T_p \ (e^{- \Phi})_{y=Z=0} \ , \label{bi12}
\ee
where  we used again the
background (\ref{bi1}) to derive the last equality.
The one-point function $g_{k,l, j}$ of the dilaton in
front of the boundary (Dp-brane) is now
\be
g_{k,l, j}= T_p (\Phi_{k,j,l})|_{y=Z=0} \ .
\label{bi13}
\ee
By using (\ref{bi8}) we find, as previously, that the non-vanishing 
wave-functions at
$Z=0$ have zero total angular momentum $l=0$. Moreover, an explicit computation
of (\ref{bi13}) finally gives the simple result 
\be
g_{k,0,j} = T_p \sqrt{\omega_k \over \pi} \ . \label{bi14}
\ee

We now turn to the string theory description of D-branes and in
particular to the brane-brane interactions which allow an independent 
check of the one-point couplings (\ref{bi11}) and (\ref{bi14}).
\section{D-Branes in the Melvin background: string amplitudes}

\subsection{D branes parallel to the Melvin space}
Let us turn to study the spectrum and interaction of D-branes in
the Melvin background and start with Neumann boundary conditions
for all relevant coordinates (y,$\rho$,$\phi$). The open string
hamiltonian describing strings stretched between two Dp-Dp branes
separated by the distance $r$, reads
\be
H = N + {\alpha' } ({k \over R}- B J)^2 + {r^2 \over 4
  \pi^2 \alpha'} +a \ , \label{d1}
\ee
where $a$ vanishes in the Ramond sector and equals $-1/2$ in the
Neveu-Schwarz sector. The usual GSO projection removes the open
string tachyon.
The spectrum of massless open string states for a stack of $M$ Dp
branes is based on the unitary gauge group $U(M)$. At the massless
level, boson masses for the $SO(p-2)$ Lorentz group are unaffected by 
the (closed) magnetic field $B$, whereas internal vectors along the
twisted plane and spacetime fermions acquire a 
mass proportional to $B^2$, thus breaking supersymmetry.
Notice that the boundary conditions do not allow the presence of
Landau levels in the open spectrum.
The quantum numbers of a state include the value of the orbital
angular momentum $m=xp_y-yp_x$ which is an integer, as well as
the norm of the momentum in the $Z_0$ plane and the components of
the momentum $q_\mu$ along the remaining D-brane coordinates.
Consider the lowest level vector states $\psi^\mu_{-1/2}|0>$  ,
then the mass shell condition reads $q^2=(k/R-Bm)^2$,
whereas for the states $\lambda_{-1/2}|0>${\footnote{remember that
$\lambda$ is
the superpartner of $Z_0$}.} it reads $p^2=(k/R-B(m\pm1))^2$.

The one-loop cylinder amplitude encoding the Dp-Dp brane interaction
is given by
\be
A_{pp} = {1 \over 2} 
\int_0^{\infty} {dt \over t} \ Tr e^{-\pi t 
(H+\alpha'q^2) } \ , \label{d2}
\ee
where $t$ is the open channel cylinder modulus and the trace
includes an integration over the momenta $q$ along the
worldvolume spanned by the Dp brane. 
As in the closed string case 
(\ref{parti}), the result is
easily written in terms of the Poisson-resummed momentum $\tilde
k$. In terms of it, the amplitude (\ref{d2}) reads
\be
A_{pp} = { R\over \sqrt{4\alpha'}} 
\sum_{\tilde k} \int_0^{\infty} {dt \over t^{3/2}} \   
e^{-{r^2 t \over 4 \alpha'\pi}} 
Tr e^{ 2 \pi i {\tilde k} B R J -  \pi t  (N+a +q^2 \alpha')} 
e^{-\pi  {\tilde k}^2 R^2/(\alpha' t) } \ , \label{d02}
\ee
 The final result of the computation, obtained by using the eqs. 
(\ref{zerom1})-(\ref{tracezero}) is
\ba
&&A_{pp} = {\pi R V_p \over{2 (4\pi^2\alpha')^{(p+1)/2}}}
 \int {dt \over t^{p+3 \over 2}} e^{-{r^2 t \over 4 \alpha'
\pi}} \sum_{\a,\b=0,1/2} \eta_{\a,\b} \
\frac{\vartheta^4[{\a \atop \b}]}{\eta^{12}} + \nonumber \\
&&{\pi R \alpha'V_{p-2} \over{( 4\pi ^2 \alpha')^{(p+1)/2}}}
\sum_{\alpha, \beta} \int {dt \over t^{p+1 \over 2}}  
e^{-{r^2 t \over 4 \pi \alpha'}} 
\sum_{\tilde k \not=0}{1 \over{\sin\pi BR\tilde k}}
\eta_{\a,\b} \frac{\vartheta^3[{\a \atop \b}] \vartheta[{\a \atop \b+
BR \tilde k}] }{\eta^{9} \vartheta[{1/2 \atop 1/2 + B R \tilde
k}]} e^{-{\pi {\tilde k}^2 R^2 \over \alpha' t}} \ , \label{d3}
\ea
where the argument of the various modular functions is 
$\tau = it /2$.
The interaction of the Dp branes with the closed string spectrum is,
as usual, transparent in the closed-string channel amplitude, written in
terms of the modulus $l = 2/t$. Notice first of all that in
this case $\tilde k$ are actually precisely the winding states $n$ of
the closed strings exchanged by the two Dp branes. The
resulting brane-brane interaction reads
\ba
&&\!\!\!\!A_{pp} = {\pi R V_p \over{2 (8\pi^2\alpha')^{(p+1)/2}}}
 \int {dl \over l^{9-p \over 2}} e^{-{r^2 \over 2
\pi \alpha' l}} \sum_{\a,\b=0,1/2} \eta_{\a,\b} \
\frac{\vartheta^4[{\a \atop \b}]}{\eta^{12}} + \nonumber \\
&&\!\!\!\!\!\!\!\!{\pi R \alpha'V_{p-2} \over{( 8\pi ^2 \alpha')^{(p+1)/2}}} 
\sum_{\alpha , \beta} 
\int {dl \over l^{9-p \over 2}}  e^{-{r^2 \over 2
\pi \alpha'l}} \sum_{n \not=0}{1 \over{2\sin\pi BRn}}
\eta_{\a,\b} \frac{\vartheta^3[{\a \atop \b}] \vartheta[{\a \atop \b
}](-iBRnl|il) }{i \eta^{9} \vartheta_1 (-iBRnl|il)}
e^{-{\pi n^2 R^2 l \over  2\alpha'}} \ . \label{d4}
\ea
This amplitude will be recovered in the next section by using the
boundary state formalism. Let us for the moment examine its low
energy limit and compare with the previous Born-Infeld result 
(\ref{bi11}). The limiting case ($r >> \sqrt{\alpha'}$, $B R << 1 $) 
is described
by decoupling the heavy (in this limit) string oscillators in
(\ref{d4}) and keeping only the closed string Landau levels coupling to the
Dp branes. Let us discuss separately the interaction of the NS-NS and
the RR closed states to the branes in this limit and check its
consistency. The NS-NS exchange is described by
\ba
&&A_{pp}^{\rm NS-NS} = A_{pp, massless}^{NS-NS}  \nonumber \\
&+& {\pi R \alpha'V_{p-2} \over{( 8\pi ^2 \alpha')^{(p+1)/2}}}
\int {dl \over l^{9-p \over 2}}  e^{-{r^2 \over 2
\pi \alpha' l}} \sum_{n \not=0}{1 \over{2\sin\pi BRn}}
{6 + 2 ch(2 \pi \nu l) \over sh (\pi \nu l)}
e^{-{\pi n^2 R^2 l \over 2 \alpha'}} \ , \label{d5}
\ea
where $\nu = BRn - [BRn]$.

In order to interpret the amplitude (\ref{d5}), it is important to
account for the reflection of closed string states on the D-brane
$|k,n,J, {\tilde J}> -> |-k,n,- {\tilde J},-J>$. The set of
states invariant under the reflection is $|0,n,J, -J >$. Notice that 
the condition
\be
J |\rm boundary>= - {\tilde J} |\rm boundary> \ , \label{d6}
\ee
allows for one-set of Landau levels of closed strings bouncing off the
brane. We therefore rediscovered at string level the vanishing of the
total angular momentum of closed states coupling to the brane, first
found from the Born-Infeld analysis of the previous section.
The Landau levels are easily identified in (\ref{d5})
by performing the expansion
\be
{6 + 2 ch(2 \pi \nu l) \over sh (\pi \nu l)} =
\sum_{j=0}^{\infty} [ \ 6 \ e^{-2 \pi \nu (j+{1 \over 2}) l} +
e^{-2 \pi \nu (j-{1 \over 2}) l} +  e^{-2 \pi \nu (j+{3 \over 2}) l} \ ]
\ . \label{d7}
\ee
The first term in (\ref{d7}), when plugged in (\ref{d5}), describes the
Landau levels of the (p+1) dim. dilaton. The second and the third term
describe the components of the metric transverse to the brane,
which can become tachyonic for small enough radius 
$R < \sqrt{2 \alpha'}$.

The long-range exchange of RR states between the Dp branes is described 
by the amplitude
\ba
&&A_{pp}^{\rm RR} = A_{pp, \rm massless}^{\rm RR}  \nonumber \\
&-& 8{\pi R \alpha'V_{p-2} \over{( 8\pi ^2 \alpha')^{(p+1)/2}}} 
\int {dl \over l^{9-p \over 2}}  e^{-{r^2 \over 2
\pi \alpha' l}} \sum_{n \not=0} {1 \over 2 \sin (\pi BRn) }
\ { ch( \pi \nu l) \over sh (\pi \nu l)} \ 
e^{-{\pi n^2 R^2 l \over 2 \alpha'}} \ . \label{d8}
\ea
The interaction of Landau levels of RR states with the D-branes is
obtained by performing the expansion
\be
{ ch(\pi \nu l) \over sh (\pi \nu l)} =
\sum_{j=0}^{\infty} [e^{-2 \pi \nu j l} +
e^{-2 \pi \nu (j+1) l} ] \ . \label{d9}
\ee
The amplitudes (\ref{d5}),(\ref{d8}) expanded in Landau levels according
to  (\ref{d7}),(\ref{d9}) allow to reconstruct the hamiltonian of
closed states interacting with the D-branes.
In order to achieve this, we use the closed-string propagator for a
canonically normalized scalar of mass $M$ and momentum $q$
\be
\Delta_c = {\pi \alpha' \over 2} \int_0^{\infty} dl \ e^{-{\pi \alpha'
l \over 2} (q_{\mu} q^{\mu}+M^2)} \ . \label{d10}
\ee
By using (\ref{d10}), we finally obtain
\ba
M^{2,\rm NS-NS} &=& N_{\rm osc} + {n^2 R^2 \over 2 \alpha'} +
(2j \pm 1 ) \nu \ ,
\nonumber \\
M^{2,\rm RR} &=& N_{\rm osc} + {n^2 R^2 \over 2 \alpha'} +  2j \ \nu \ ,
\label{d11}
\ea
in agreement with the argument presented in (\ref{d6}). Notice in
(\ref{d11}) the coupling of the would-be tachyon $j=0$ in the NS-NS
sector of the D-brane, corresponding to the fluctuations of the metric
transverse to the brane. 
We are now able to extract the one-point couplings of the various
closed fields and compare them with the field-theory result (\ref{bi11}).
In order to do so, notice first that by performing a T-duality on $y$
the windings $n$ become Kaluza-Klein states $k$.
Consider moreover the field-theory  limit in
(\ref{d5}), (\ref{d8}), in which (after T-duality)
$sin (\pi BRk) \simeq \pi BRk$. The string amplitudes
(\ref{d5}), (\ref{d8}) contain the tree-level propagation of closed
states between the branes and therefore contain two one-point
functions and a closed propagator. By a quick inspection we precisely
recover the field-theory result for the one-point coupling
(\ref{bi11}). It is very interesting to go beyond the field-theory
limit in (\ref{d5}), (\ref{d8}). In this case, the one-point functions
show a deviation from the field-theory result which can be attributed
to $\alpha'$ corrections to the Born-Infeld action.
This point could bring new insights into the higher derivative corrections
to the Born-Infeld action and clearly deserves a more dedicated 
investigation. 

\subsection{D branes at the origin of the Melvin space}

Here we consider the case where $y$ is parallel to the brane which
has now Dirichlet boundary conditions in the $\rho,\phi$ 
plane and moreover is at the origin of the plane, anticipated in section
2.2e. The difference compared
to the case where all relevant coordinates were Neumann is that
there are  no zero-mode (non-compact) momenta contributions from
the the $Z_0$ coordinate and the
the angular momentum has no zero mode contribution.
The final amplitude taking into account these changes is
\ba
&&A_{pp} = {\pi R V_p \over{2 (4\pi^2\alpha')^{(p+1)/2}}}
 \int {dt \over t^{p+3 \over 2}} e^{-{r^2 t \over 4 \alpha'
\pi}} \sum_{\a,\b=0,1/2} \eta_{\a,\b} \
\frac{\vartheta^4[{\a \atop \b}]}{\eta^{12}} + \nonumber \\
&&{\pi R V_{p} \over{( 4\pi ^2 \alpha')^{(p+1)/2}}}
\sum_{\alpha, \beta} \int {dt \over t^{p+3 \over 2}}  
e^{-{r^2 t \over 4 \pi \alpha'}} 
\sum_{\tilde k \not=0}{{\sin\pi BR\tilde k}}
\eta_{\a,\b} \frac{\vartheta^3[{\a \atop \b}] \vartheta[{\a \atop \b+
BR \tilde k}] }{\eta^{9} \vartheta[{1/2 \atop 1/2 + B R \tilde
k}]}
e^{-{\pi {\tilde k}^2 R^2 \over \alpha' t}} \ . \label{d12}
\ea

The identification of the tree-level closed string 
exchange between
two such D-branes follows closely the one described in 
the previous
paragraph. The only difference is in the zero-mode part, which in
the closed channel amounts to replace the factor
$2 \sin (\pi \nu) $ in various amplitudes by its inverse
$1 / (2 \sin (\pi \nu))$. 
In the low-energy limit $\nu << 1$, the amplitude containing the
square of the one-point function turns out to be very similar to that
of the NN D-branes, with the notable difference that the factor
$\omega_k$ is now replaced by $1/ \omega_k$. 
This matches the field
theory result (\ref{bi14}), however the phase factor $(-1)^j$
that was found there squares to one and cannot therefore be detected
by the present computation.

In order to resolve this ambiguity,
we now compute the cylinder amplitude between one Dp brane with
NN boundary conditions along the $Z$ plane and one D(p-2) brane
with DD boundary conditions, respectively. In the long-range,
field-theory limit, this amplitude encodes the cross product of
one-point functions of the closed fields with the two types of branes.
The resulting amplitude can be easily written by quantizing properly
the open strings with Neumann-Dirichlet boundary conditions along
$Z$. The result is most conveniently written in the closed channel
and read
\ba
&&A_{p,p-2} = {\pi R V_{p-2}
 \over{2 (4\pi^2\alpha')^{(p-1)/2}}} 
 \int {dl \over l^{9-p \over 2}} e^{-{r^2 \over 2
\pi \alpha' l}} \sum_{\a,\b=0,1/2} \eta_{\a,\b} \
\frac{\vartheta^3[{\a \atop \b}] \vartheta[{\a \atop \b+1/2}] }
{\eta^{9} \vartheta_2} + \nonumber \\
&& {\pi R V_{p-2}
 \over{2 (4\pi^2\alpha')^{(p-1)/2}}}\sum_{\alpha , \beta} 
\int {dl \over l^{7-p \over 2}}  e^{-{r^2 \over 2
\pi \alpha'l}} \sum_{n \not=0}
\eta_{\a,\b} \frac{\vartheta^3[{\a \atop \b}] \vartheta[{\a \atop \b
+1/2}](-iBRnl|il) }{i \eta^{9}  \vartheta_2 (-iBRnl|il)}
e^{-{\pi n^2 R^2 l \over 2 \alpha'}} \ . \label{d13}
\ea

The long-range exchange of RR states between the Dp and the D(p-2)
brane turns out to be now given by
the amplitude
\ba
&&A_{p,p-2}^{\rm RR} = A_{p,p-2, \rm massless}^{\rm RR}  \nonumber \\
&-& 8 {\pi R V_{p-2} \over{(4\pi^2\alpha')^{(p-1)/2}}}  
\int {dl \over l^{7-p \over 2}}  e^{-{r^2 \over 2
\pi \alpha' l}} \sum_{n \not=0}
{ sh( \pi \nu l) \over ch (\pi \nu l)}
\ e^{-{\pi n^2 R^2 l \over 2 \alpha'}} \ . \label{d14}
\ea
The interaction of Landau levels of RR states with the D-branes is
now obtained by performing the expansion
\be
{ sh(\pi \nu l) \over ch (\pi \nu l)} =
\sum_{j=0}^{\infty} (-1)^j [e^{-2 \pi \nu j l} -
e^{-2 \pi \nu (j+1) l} ] \ . \label{d15}
\ee
Plugging in (\ref{d15}) in (\ref{d14}) we finally obtain the cross
product of one-point functions of RR fields with the two different
D-branes. The presence of the phase factor $(-1)^j$ and the
absence of the zero mode $\sin(\pi \nu)$ factor are precisely required
for the exact agreement in the field theory limit $\nu << 1$ of the
one-point functions derived from (\ref{d15}) with the ones derived
from the Born-Infeld action. Similar results hold for the couplings
of the NS-NS fields to the branes.

\subsection{D branes with fixed $Z_0$}

We consider here D-branes parallel to $y$ but perpendicular to the
$(\rho,\phi_0)$ plane.
Here $y$ is a coordinate along the brane which has a fixed 
but nonzero value for $Z_0$. As explained in section 2.2.d
the $y$ coordinate is a quasi-periodic coordinate.
The cylinder amplitude is given by
\be
A_{pp}={V_p \over {4 (4\pi^2\alpha')^{p/2}}}
\int {dt \over t^{p+2 \over 2}}\sum_{k,l} e^{-\pi
t\alpha'(k/R-Bl)^2} \sum_{\alpha,\beta} \eta_{\alpha \beta}
{\vartheta^4[^\alpha_{\beta}] \over \eta^{12}} \ ,
\ee
where the sum is the contribution of the momenta of the $y$
coordinate. Interestingly enough, the partition function  describes a 
supersymmetric D-brane open spectrum.

The Kaluza-Klein contribution of the $y$ momenta can
be put in the form
\be
\sum_{k,l} e^{-\pi
t\alpha'(k/R-Bl)^2}=\sum_{l} \vartheta[^{RBl}_{0}]
(0|{it\alpha' \over R^2}) \ . \label{som}
\ee
A modular transformation allows the sum to be cast in the form
\be
{R \over \sqrt{t\alpha'}}\sum_{l,\tilde k}e^{-{\pi R^2 \tilde k^2
\over \alpha' t}} e^{2\pi i RB l \tilde k} \ .
\ee
The sum is divergent as it stands, therefore a usual 
finite volume regularisation is required.
The simplest finite volume regularisation is to assume that
$BR$ is rational of the form
$P/ Q$ where $P$ and $Q$ are large integers with 
no common divisor.
The non-compact space $Y$ space is replaced with a circle of radius $L=QR$.
The sum (\ref{som}) is replaced with
\be
\sum_{k \in Z}\sum_{l=0}^{Q-1} e^{-\pi t\alpha'(k-Pl /Q)^2/R^2}=
{R \over \sqrt{t\alpha'}} \sum_{\tilde k} e^{-{\pi R^2 \tilde k^2
\over \alpha' t}} \sum_{l=0}^{Q-1} e^{2\pi i Pl \tilde k/Q} \ .
\ee
The sum over $l$ can be readily calculated and gives
$Q\sum_p\delta_{\tilde k,pQ}$ which allows the regularised 
sum (\ref{som}), in the limit of large $L$,
to be cast in the form
\be
{L \over \sqrt{t\alpha'}} \ . \label{som2}
\ee

\subsection{D branes with fixed $y$ and $\phi$}

Here we consider open strings with Dirichlet boundary conditions 
for $y$ and $\phi_0$ or equivalently for $y$ and $\phi$. The
coordinate $\rho$ is taken to be Neumann, such that the brane
intersects the $\rho, \phi$ plane with a line of constant $\phi$.
The bosonic Hamiltonian was presented 
in subsection 2.2b, where it was shown that the contribution
from the $Z$ coordinates is that of a twisted boson. 
The fermionic hamiltonian  can be deduced similarly,
since the superpartners of $Z_0$ have the same twisting
in the Ramond sector and  an opposite one in the Neveu-Schwarz sector.
The hamiltonian can be put in the form
\be
H=N-\nu J+a \ ,
\ee
where $N$ is as usual the number operator, $J$ 
is the total  angular momentum and $a$ vanishes in the Ramond
sector and is given by
$(\nu-1)/2$ in the Neveu-Schwarz sector.
The cylinder amplitude in the direct channel
encoding the GSO projection  is given by 
\ba
&&A_{pp} =  {V_{p+1} \over{2 (4\pi^2\alpha')^{(p+1)/2}}}
 \int {dt \over t^{p+3 \over 2}} e^{-{r^2 t \over 4
\pi\alpha'}} \sum_{\a,\b=0,1/2} \eta_{\a,\b} \
\frac{\vartheta^4[{\a \atop \b}]}{\eta^{12}}  \nonumber \\
&&+ {i V_{p} \over{2 (4\pi^2\alpha')^{p/2}}}  \sum_{\alpha , \beta} 
\int {dt \over t^{p+2 \over 2}}  e^{-{r^2 t \over 4
\pi\alpha'}} \sum_{n \not=0}
\eta_{\a,\b} \frac{\vartheta^3[{\a \atop \b}]
\vartheta[{\a \atop \b}] (iBRnt|it/2)}{\eta^{9}
\vartheta_1 (iBRnt|it/2)}
e^{-{\pi {n}^2 R^2 t}/\alpha'} \ . \label{d33}
\ea
Notice that this amplitude can be constructed starting from the
cylinder amplitude for a D-brane in flat space
and then adding the images 
which arise upon the identification (\ref{one}).
The interaction of a brane with its image is given by the
amplitude
of two branes distant in the $y$ direction by $n2\pi R$ and at a
relative angle
$2nBR\pi$.
In order to verify the consistency of the amplitude we have to 
consider its closed-string interpretation:
\ba
&&A_{pp} = {V_{p+1} \over{2 (8\pi^2\alpha')^{(p+1)/2}}} 
\int {dl \over l^{9-p \over 2}} e^{-{r^2 \over 2
\pi \alpha' l}} \sum_{\a,\b} \eta_{\a,\b} \
\frac{\vartheta^4[{\a \atop \b}]}{\eta^{12}} + \nonumber \\
&\!\!+\!\!& {V_{p} \over{2 (8\pi^2\alpha')^{p/2}}}
\int {dl \over l^{8-p \over 2}}  
e^{-{r^2 \over 2 \pi \alpha' l}} \sum_{\tilde k \not=0}
\sum_{\a,\b} \eta_{\a,\b} \ \frac{\vartheta^3[{\a \atop \b}]
\vartheta[{\a \atop \b+2RB\tilde k}]}{\eta^{10}}
\frac{\eta}{\vartheta[{1/2 \atop 1/2+2RB\tilde
k}]}e^{-2\pi\tilde k^2 R^2/l} \ . \label{tr2}
\ea
This is the form of the amplitude that we must recover in the 
next section from the closed string tree level propagation of 
the boundary states. 
  Notice that, contrary to the examples discussed in sections 4.1 and
4.2, here the open sector do contain Landau levels, while the closed states
interacting with the brane do not, due to the reflection on the brane. 

\subsection{D branes transverse to the Melvin space}

Here the three coordinates $y$ , $\rho$ and $\phi$ are Dirichlet
coordinates. From the discussion in section 2.2.c,
the partition function can be readily determined: 
\be
A_{pp} = {V_{p+1} \over 2 (4\pi^2\alpha')^{(p+1)/2}}
 \sum_{n}\int {dt \over t^{p+3 \over 2}}
e^{-{r^2+4\rho_0^2 sin^2(\pi BRn)+(2\pi nR)^2 \over 4
\pi \alpha'} t } \sum_{\a,\b=0,1/2} \eta_{\a,\b} \
\frac{\vartheta^4[{\a \atop \b}]}{\eta^{12}} \ . \label{quatre5}
\ee
The open spectrum is supersymmetric, and the sine squared
term contribution to the mass is due to the fact that 
if the values of $Z$ at the two ends of the open string coincide
then the values of $Z_0$ do not. 
\section{Boundary states}

The one loop open channel amplitudes
are constrained by the requirement of their dual
interpretation as tree level closed strings exchange.
The boundary states $|B>$ encode this interpretation via
\cite{boundary}:
\be
A_{pp}=\int dl <B|e^{-\pi l(L_0+\bar L_0)}|B> \ ,
\ee
where $l$ is the closed string modular parameter related to the
open one by $l=2/t$.
In the following we shall determine the boundary states 
and verify the correct factorisation of the one loop amplitudes.

\subsection{The $y$ coordinate is Neumann }

The boundary state verifies
\be
\partial_{\tau}y(0,\sigma)|B,\eta>=0 \ , \ (\psi_+(0,\sigma)-\eta
\psi_-(0,\sigma))|B,\eta>=0 \ ,
\ee
where $\psi$ is the worldsheet superpartner of $y$.
In terms of the mode expansion we get
\ba
&&\Big({k \over R}-B(J+\tilde J)\Big)|B,\eta>=0 \ , \ 
(y_m-\tilde y_m^\dagger) |B,\eta>=
(y^\dagger_m-\tilde y_m)|B,\eta>=0 \ , \\
&& (\psi_m-\eta\tilde \psi_{-m})|B,\eta>=0 \ .
\ea
Since $BR$ is irrational we get from the first equation that both
$k$ and $J+\tilde J$ must vanish.
The boundary state has the form
\be
|B,\eta>=\sum_n N_n\ |n>\otimes \exp\{\sum_{m=1}^{\infty}
({y_m}^\dagger{\tilde y_m}^\dagger+\eta \psi_{-m}\tilde
\psi_{-m})\}|0>\otimes|\psi_n,\eta> \ ,
\ee
where $n$ denotes the winding mode, $N_n$ is a normalisation
constant, and $|\psi_n,\eta>$ is the state depending on the other
coordinates which  has to verify
\be
(J+\tilde J)|\psi_n,\eta>=0 \ .\label{contra}
\ee
As usual, the combination which is correctly GSO projected is
$|B>=|B,+>+|B,->$.
Since the world-sheet fermions contribution has minor
modifications with respect to the flat space one,
in the following we shall concentrate on the 
nontrivial bosonic contribution which is due
to the coordinates $y$
and $Z_0$. 

{\bf a}) $Z_0$ Neumann. The state $|\psi_n,\eta>$ verifies
\ba
&&(a_m-\tilde b_m^{\dagger})|\psi_n,\eta>=0 \ , \
(\tilde a_m- b_m^{\dagger})|\psi_n,\eta>=0 \ , \nonumber\\
&&(a_m^\dagger-\tilde b_m)|\psi_n,\eta>=0 \ , \
(\tilde a_m^\dagger- b_m)|\psi_n,\eta>=0 \ . \nonumber\\
&&(\lambda_m- \eta\tilde \lambda_{-m})|\psi_n,\eta>=0 \ .
\ea
When $n$ is zero, the state $|\psi_0,\eta>$ has zero
momentum in the $Z_0$ directions.
The state $|\psi_n,\eta>$ is thus given by
\be
|\psi_n,\eta> = \exp\{\sum_{m=0}^\infty \tilde a_m^\dagger b_m^\dagger
+\sum_{m=1}^\infty  a_m^\dagger \tilde b_m^\dagger\}|0> \ .
\ee
This automatically verifies the constraint (\ref{contra}).
The normalisation constants can be determined from the
comparison with the open channel amplitudes, and the result is
$N_n^2={N \over \sin \pi BRn}$.

{\bf b}) $Z_0$ Dirichlet. The state $|\psi_n,\eta>$ has to
satisfy the constraint
\be
(Z_0(\tau=0,\sigma)-z_0)|\psi_n,\eta>=0 \ .
\ee

The treatment of the nonzero modes is similar
to the previous  case, that is
\ba
&&(a_m+\tilde b_m^{\dagger})|\psi_n,\eta>=0 \ , \
(\tilde a_m+ b_m^{\dagger})|\psi_n,\eta>=0 \ , \nonumber\\
&&(a_m^\dagger+\tilde b_m)|\psi_n,\eta>=0 \ , \
(\tilde a_m^\dagger+ b_m)|\psi_n,\eta>=0 \ . \nonumber\\
&&(\lambda_m+\eta\tilde \lambda_{-m})|\psi_n,\eta>=0 \ .
\ea
The main difference comes from the equations of the zero modes
which depend crucially on whether $z_0$ vanishes or not.
In fact when $n \neq 0$, $Z_0$ has no zero mode part
and so if $z_0 \neq 0$ then  necessarily we have
\be
N_n=0 \ , n\neq 0 \ .
\ee
This agrees with the form of the open channel amplitude (\ref{som2})
found in section 4.3 where it was shown that in the transverse
channel there are no closed string winding  or Kaluza-Klein 
modes propagating in the amplitude.

When $z_0=0$, the normalisation constants are determined from
the open channel amplitudes
found in section 4.2. We get $N_n^2=N \sin \pi BRn$.

{\bf c}) $X_0$ Dirichlet and $Y_0$ Neumann.
There is a contradiction between (\ref{contra}) and the
conditions expressing different boundary conditions for 
$X_0$ and $Y_0$. This rules out this case as a consistent
D-brane configuration.

\subsection{The $y$ coordinate is  Dirichlet}

Here the boundary state must verify
\be
\Big(y(0,\sigma)-y_0\Big)|B,\eta>=0 \ , \ (\psi_+(0,\sigma)+\eta
\psi_-(0,\sigma))|B,\eta>=0 \ , \label{zer}
\ee
which imply that there are no winding modes contributing to the
boundary state and $|B,\eta>$ has the form
\be
|B,\eta>=\sum_k N e^{iky_0/R} |k>\otimes \exp\{-\sum_{m=1}^{\infty}
({y_m}^\dagger{\tilde y_m}^\dagger + \eta \psi_{-m}\tilde
\psi_{-m}) \}|0>\otimes|\psi_k,\eta> \ ,
\ee
where $k$ is the Kaluza-Klein mode. 
Since the winding mode is zero there is no
twisting of the $Z_0$ coordinate or its superpartner.
There is no constraint analogous to (\ref{contra}) so it seems
that all boundary conditions for $Z_0$ are consistent. 
This is the main difference with the respect
to the case where the coordinate $y$ is Neumann.
 
{\bf a}) $Z_0$ Neumann.
Here the boundary state has zero momentum along the $Z_0$
direction and 
$(J+\tilde J)|B,\eta>=0$, so the closed string hamiltonian
acting on the boundary state is the same as the free one.
This agrees with the results of section 2.2f. 
  
{\bf b}) $Z$ Dirichlet, which implies also that $Z_0$ is Dirichlet.
However, if $Z$ has the same value at the two ends of the open string
then this is not the case for $Z_0$ in a $y$ winding sector. 
This gave rise in section 4.5 to a nontrivial 
factor $\rho_0^2\sin^2(\pi
BRn)$ in the amplitude (\ref{quatre5}). In order to retrieve
the origin of this factor in the closed string
let us remark that the oscillator contribution to $J+\tilde J$
vanishes on the boundary state, so it is consistent to focus on
the zero modes contribution of the amplitude. The
zero mode part of the boundary state reads
\be
|b>=N\sum_k\int d^2p \ e^{i{\bf p.r_0}} \ |{\bf p},k> \ ,
\ee
where ${\bf r_0}$ denotes the position of the brane in the $Z_0$
plane, with $|{\bf r_0}|=\rho_0$
and $N$ is a normalisation constant. 
The zero mode contribution to
the amplitude reads
\be
<b|e^{-\pi l\alpha'[(p_y-BL)^2+{\bf p}^2]/2}|b> \ ,
\ee 
where $L$ is the orbital momentum operator in the $Z_0$ plane.
 A Poisson resummation on the Kaluza-Klein modes allows this
 amplitude to have the form
 \be
 N^2{R \sqrt{2}\over \sqrt{l\alpha'}}
 \sum_n\int d^2p_1 d^2p_2 e^{-2\pi R^2n^2/(\alpha'l)}
 <{\bf p_1}|e^{i2\pi L RBn}|{\bf p_2}> e^{-\pi l \alpha' {\bf
 p_2/2}^2}e^{-i({\bf p_1}-{\bf p_2}).{\bf r_0}} \ .
 \ee
 Let $\theta$ denote the rotation in the $Z_0$ plane with an
 angle $-2\pi BRn$ then the integral can be readily calculated to
 give
 \be
 N^2{R2\sqrt{2} \over \sqrt{(l\alpha')^3}}
 e^{-2 \pi R^2n^2/(\alpha'l)}
 e^{-(\bf r_0-\theta \bf r_0)^2/(2\pi\alpha' l)}=
 N^2{R 2\sqrt{2}\over \sqrt{(l\alpha')^3}}
 e^{-2\pi R^2n^2/(\alpha'l)}
 e^{-2\rho_0^2\sin^2(\pi BRn)/(\pi
 l\alpha')} \ . \label{b1}
 \ee
 By using the open modulus $t = (2/l)$ we find that (\ref{b1}) 
 is in perfect agreement with the open channel amplitude 
 (\ref{quatre5}) and so the consistency of
 the latter is nicely verified.

{\bf c}) $\phi$ Dirichlet and $\rho$ Neumann.
Since $\phi_0$ in the direct channel satisfies a Dirichlet
boundary condition,
the D-brane boundary state verifies $\partial_\sigma\phi_0|B>=0$
which leads to $(J-\tilde J)|B>=0$. When acting on the boundary
state the closed string Hamiltonian reduces to 
$H=H_0+ (\alpha'/2) (k/R-B(J+\tilde J))^2$ and therefore
\be
<B|e^{-\pi l(L_0+\bar L_0)}|B>=\sum_{k}
<B|e^{-\pi l (H_0+\alpha'/2(k/R-B(J+\tilde J))^2)}|B> \ .
\ee
Performing a Poisson resummation we get
\be
<B|e^{-\pi l(L_0+\bar L_0)}|B>= \sum_{\tilde k}\sqrt{R^2\alpha' \over l}
<B|e^{-\pi l H_0} e^{\pi i 2 BR(J+\tilde J) \tilde k}|B>
e^{-{2\pi\tilde k^2 R^2\alpha'\over l}} \ . \label{bo1}
\ee
By using the form of the boundary states we can explicitly
evaluate (\ref{bo1}). In fact, the dependence of $|B>$
on the $Z_0$ coordinate is given by
\be
|b> \times \exp\{-\sum_{m=1}^{\infty} (a_m^\dagger\tilde a_m^\dagger
+b_m^\dagger\tilde b_m^\dagger) \} |0> \ , \label{bo2}
\ee
where $|b>$ is the zero modes contribution which
reads
\be
|b>=\int dp_{y_0} \ |p_{x_0}=0,p_{y_0}> \ , \label{bo3} 
\ee
where we have assumed for simplicity that $\phi_0=0$
on the boundary of the open string.
The bosonic contribution in the $(\rho,\phi_0)$ 
plane  in (\ref{bo1}), computed by using (\ref{bo2})-(\ref{bo3}) 
and performing some standard computations turns out to be equal to
\be
 {e^{ \pi l \over 6} \over  \sin (2\pi BR{\tilde k})}
\prod_{m=1}^{\infty} {1 \over 
(1-e^{4 \pi i BR{\tilde k}} e^{-2 \pi lm})
 (1-e^{-4 \pi i BR{\tilde k}} e^{-2 \pi lm})} = 
{\eta \over \vartheta [{1/2 \atop 1/2+ 2BR{\tilde k} }]}
(q = e^{-2 \pi l}) \ , \label{bo4}
\ee
where the $\sin (2\pi BR{\tilde k})$ factor 
comes from the zero modes contribution .
Collecting together the bosonic and fermionic contributions, 
we find indeed agreement with (\ref{tr2}).
\section{Supersymmetric Melvin solutions}

Supersymmetric Melvin solutions, preserving one-half supersymmetry, 
are easily obtained by accompanying the
complete translation $y \rightarrow y + 2 \pi R$ around the circle 
with correlated
rotations in the noncompact planes $(\rho_1,\phi_1)$,  
$(\rho_2,\phi_2)$. More precisely, the magnetic field parameters
corresponding to the two planes are $B_1 = \pm B_2$.
The metric of the resulting models is
\be
ds^2=d\rho_1^2+d\rho_2^2+ 
\rho_1^2(d\phi_1+B dy)^2+ \rho_2^2(d\phi_1 \pm B dy)^2
+dy^2+d{\bf x}^2 \ , \label{s1}
\ee
where ${\bf x}$ denote here the resulting 5 space-time coordinates.

The torus amplitude for type IIB (IIA)
in the Melvin background is given by
\be
T=V_5 {R\over \sqrt{\pi\alpha'\tau_2}}
{1 \over(4 \pi^2\alpha'\tau_2)^{5/2}|\eta|^{8}}
\Big\{\sum_{\tilde k, n}e^{-{\pi R^2 \over \alpha'\tau_2}
|\tilde k+\tau n|^2}Z(\tilde k,n)\Big\},\label{s2}
\ee
where
\ba
Z(\tilde k,n)&=& {1 \over 4} | \sum_{\alpha,\beta}\eta_{\alpha
\beta}{\vartheta^2[^\alpha_\beta] \over \eta^2}
{\vartheta[^{\alpha+RBn}_{\beta+RB\tilde k}]
\vartheta[^{\alpha-RBn}_{\beta-RB\tilde k}]
\over \vartheta[^{1/2+RBn}_{1/2+RB\tilde k}] 
\vartheta[^{1/2-RBn}_{1/2-RB\tilde k}]} |^2 \ ,
\nonumber \\
Z(0,0) &=& {1 \over 4} {V_4 \over (4\pi^2\alpha'\tau_2)^2|\eta|^8}
\Big\{\sum_{\alpha,\beta}\eta_{\alpha
\beta}{\vartheta^4[^\alpha_\beta] \over \eta^4}
\Big\}
\Big\{\sum_{\alpha,\beta}\bar \eta_{\alpha
\beta}{\bar \vartheta^4[^\alpha_\beta] \over \bar \eta^4}
\Big\} \ .
\label{s3}
\ea
Using a Jacobi identity it is possible to cast
(\ref{parti}) in the Green-Schwarz form
\be
Z(\tilde k,n)=\Big|{\vartheta_1^2 \vartheta_1^2(BR(\tilde k+n\tau)/2|\tau)
\over \eta^3 \vartheta_1(BR(\tilde k+n\tau)|\tau)}\Big|^2 \ , \label{s4}
\ee
which clearly exhibits the resulting one-half supersymmetry of the
closed string spectrum.
The spectrum and interaction of D-branes in this supersymmetric
background is a straightforward generalisation of those already
presented in section 4.1. For example, in the case where all (five)
relevant coordinates of this background are parallel to the D-branes
under consideration, the generalization of (\ref{d3}) to this case
is
\ba
\!\!\!\!\!\!&&A_{pp} = {\pi R V_p \over{2 (4\pi^2\alpha')^{(p+1)/2}}}
 \int {dt \over t^{p+3 \over 2}} e^{-{r^2 t \over 4 \alpha'
\pi}} \sum_{\a,\b=0,1/2} \eta_{\a,\b} \
\frac{\vartheta^4[{\a \atop \b}]}{\eta^{12}} + \nonumber \\
&&\!\!\!\!\!\!{\pi R {\alpha'}^2 V_{p-4} \over{2 ( 4\pi ^2 \alpha')^{(p+1)/2}}}
\sum_{\alpha, \beta} \int {dt \over t^{p-1 \over 2}}  
e^{-{r^2 t \over 4
\pi \alpha'}} 
\sum_{\tilde k \not=0}{1 \over{(2\sin\pi BR\tilde k)^2}}
\eta_{\a,\b} \frac{\vartheta^2[{\a \atop \b}] \vartheta^2[{\a \atop \b+
BR \tilde k}] }{\eta^{6} \vartheta^2[{1/2 \atop 1/2 + 
B R \tilde k}]}
e^{-{\pi {\tilde k}^2 R^2 \over \alpha' t}} \ . \label{s5}
\ea
It is easy to see that, compared to the absence of the magnetic field
parameters, half of the supersymmetry is broken on the D-branes.
At the massless level, however, the full original supersymmetry is
restored.  

\section{The case of rational magnetic field}

Previously the twist $BR$ was supposed to be irrational.
Here, we briefly examine the case of a rational twist.
We consider, for simplicity the case where
\be
BR={1 \over N}.
\ee
The identification (\ref{one}) reads
\be
(y,\phi_0)=(y+2\pi n_1,\phi_0+{2\pi n_1 \over N}+2\pi n_2).
\ee
An equivalent description is to start with $y$ describing
a circle of radius $NR$, that is
\be
y=y+2\pi RN,
\ee
and then modd out by the $Z_{N}$ transformations
\be
(y,\phi_0)=(y+n_1 2\pi R,\phi_0+{2\pi n_1 \over N}), \
n_1=1,\dots N-1.
\ee
Notice that this a freely acting orbifold with no fixed points.
The torus partition function reads
\be
T=V_7{R\over \sqrt{\pi\alpha'\tau_2}}{1 \over (4\pi^2\alpha' 
\tau_2)^{7/2}|\eta|^{12}} {1 \over N}
\Big\{\sum_{\tilde K\in Z, {\bar N} \in Z}
\sum_{\tilde k, n=0}^{N-1}
e^{-{\pi R^2 \over \alpha'\tau_2}
|\tilde K N+\tilde k+\tau ({\bar N} N+n)|^2}
Z(\tilde k,n)\Big\} \ , \label{tor3}
\ee
with
\ba
Z(\tilde k,n)&\!\!\!=\!\!\!&{1 \over 4} \Big\{\sum_{\alpha,\beta}\eta_{\alpha
\beta} e^{2\pi i (\alpha {\tilde K} - \beta {\bar N} - \beta {n\over N})}
{\vartheta^3[^\alpha_\beta]
\vartheta[^{\alpha+{n \over N}}_{\beta+{\tilde k\over N}}] \over 
\eta^3 \vartheta[^{1/2+{n\over N}}_{1/2+{\tilde k\over N}}]}
\Big\} \Big\{\sum_{\alpha,\beta}\bar \eta_{\alpha
\beta} e^{-2\pi i (\alpha {\tilde K} - \beta {\bar N} - \beta {n\over N})}
{\bar \vartheta^3[^\alpha_\beta]
\bar \vartheta[^{\alpha+{n\over N}}_{\beta+{\tilde k\over N}}] 
\over \bar \eta^3 \bar\vartheta[^{1/2+{n\over N}}_{1/2+{\tilde k\over N}}]}
\Big\} \ , \nonumber \\
Z(0,0)&\!\!=\!\!&{V_2 \over 16 \pi^2\alpha'\tau_2} {1 \over |\eta|^4}
\Big\{\sum_{\alpha,\beta}\eta_{\alpha
\beta}{\vartheta^4[^\alpha_\beta] \over \eta^4}
\Big\}
\Big\{\sum_{\alpha,\beta}\bar \eta_{\alpha
\beta}{\bar \vartheta^4[^\alpha_\beta] \over \bar \eta^4}
\Big\} \ .
\label{partit}
\ea
In the $R \rightarrow \infty$ limit, this reduces to the 10d torus
type II partition function, whereas in the $R \rightarrow 0$ limit
it reduces to the $C/Z_N$ non-compact orbifold. 
Notice the similarity with the generalisation to string theory
of the Scherk-Schwarz \cite{ss} partition function found in 
\cite{ssclosed}. There are however two important differences:
$N$ is arbitrary, whereas in \cite{ssclosed} $N$ can take only a very
limited set of values and the two-dimensional space $(\rho,\phi)$
is non-compact.

\subsection{Open strings}
The rational magnetic case, equivalent to noncompact freely-acting 
orbifolds, have D-brane spectra that generalize the ones worked
out in the compact Scherk-Schwarz case \cite{ssopen}. For example, if all
coordinates are Neumann, we expect deformations
of the D-brane spectra with masses proportional to $1(NR)$. 
On the other hand, due to the interpolation induced by the radius $R$
in this case, the D-branes interpolate between 10d D-brane
spectra and D-branes in $C/Z_N$ non-compact orbifolds. 
 
An important difference with respect to the irrational case is
that here it is possible to consistently have D-branes which
were inconsistent in the irrational twist case.
Consider, for instance, the case with Neumann boundary conditions
for $y$ and $\rho$ and Dirichlet for $\phi_0$.
To implement it,  it is convenient to use the equivalent description 
of the background as a freely acting orbifold of a flat spacetime 
with a circle of
radius $NR$. We start with the radius $NR$ and Dirichlet
condition for $\phi_0$, then we add the $N-1$ images
accompagnied by the shift in the $y$ coordinate.
The resulting cylinder amplitude reads
\ba
&&{V_p \over {2 N (4\pi^2\alpha')^{p/2}}}\int {dt \over {t^{p+1 \over
2}}}\Big[ \sum_k e^{-{\pi t\alpha'k^2 \over N^2R^2}}
\sum_{\alpha,\beta}\eta_{\alpha\beta}
{\vartheta^4[^\alpha_\beta] \over \eta^{12}}\nonumber\\
&&+ \sum_{n=1}^{N-1}\sum_k e^{2 i \pi k n \over N}
e^{-{\pi t\alpha'k^2 \over N^2R^2}}
\sum_{\alpha,\beta} \eta_{\alpha\beta} \ 
e^{-2\pi i (\beta-{1 \over 2}) {n \over N}}
{\vartheta^3[^\alpha_\beta] \vartheta[^{\alpha+{n \over N}}_\beta] 
\over \eta^{10}}
{\eta \over  \vartheta[^{{1 \over 2}+{n \over N}}_{1 \over
2}]}\Big] \ .
\ea
Using the equivalence with the freely acting orbifold it is 
straightforward to examine the other D-branes in this background.
\vskip 1cm
\noindent
{\bf Acknowledgments:} We are grateful to C. Angelantonj
for collaboration at the beginning of this project. 
E.D. would like to thank the Theory Group at LBNL-Berkeley for hospitality
during the final stage of this work. Work supported in part
by the RTN European Program HPRN-CT-2000-00148. 

\appendix

\section{Jacobi functions and their properties}

For the reader's convenience we collect in this Appendix the
definitions, transformation properties and some identities 
among the
modular functions that are used in the text. 
The Dedekind function is
defined by the usual product formula (with $q=e^{2\pi i\tau}$)
\be
\eta(\tau) = q^{1\over 24} \prod_{n=1}^\infty (1-q^n)\ ,
\ee
whereas the Jacobi $\vartheta$-functions with general 
characteristic and
arguments  are
\ba
&&\vartheta \left[{\alpha \atop \beta }\right] (z|\tau) = \sum_{n\in Z}
e^{i\pi\tau(n-\a)^2} e^{2\pi i(z- \b)(n-\a)} \\
&&=q^{\alpha^2/2} e^{2\pi i \alpha(\beta-z)}
\prod_{m=1}^\infty(1-q^m)(1+e^{-2\pi i (z-\b)}q^{m+\a-1/2})
 (1+e^{2\pi i (z-\b)} q^{m-\a-1/2}) \ . \nonumber
\ea
In the text we have used the definition 
$\vartheta_1(z|\tau)=\vartheta[{1/2 \atop 1/2}](z|\tau)$.
The modular properties of these functions are described by
\be
\eta(\tau+1) = e^{i\pi/12}\eta(\tau)\ \ , \ \
\vartheta \left[{\a \atop {\b}}\right] \left({z} |
  {\tau+1}\right)=
e^{-i\pi\a(\a-1)}\vartheta 
\left[{\a \atop {\a+\b-{1\over 2}}}\right] \left({z} \ |
  {\tau}\right) \ , \nonumber 
\ee
\be
\eta(-1/\tau) = \sqrt{-i\tau}\; \eta(\tau)\ \ , \ \ 
\vartheta \left[{\a \atop {\b}}\right] 
\left({z \over \tau}| {-1 \over \tau}\right)=
\sqrt{-i \tau} \ e^{2 i \pi \a \b +{i \pi z^2 / \tau}} \ 
\vartheta \left[{{\b} \atop - \a}\right] (z| \tau ) \ . \label{f8}
\ee



\begin{thebibliography}{999}

\bibitem{Dai:1989ua}
J.~Dai, R.~G.~Leigh and J.~Polchinski,
Mod.\ Phys.\ Lett.\ A {\bf 4}, 2073 (1989).

\bibitem{Polchinski:1995mt}
J.~Polchinski,
Phys.\ Rev.\ Lett.\  {\bf 75}, 4724 (1995)
[hep-th/9510017].

\bibitem{Douglas:1997yp}
M.~R.~Douglas, D.~Kabat, P.~Pouliot and S.~H.~Shenker,
Nucl.\ Phys.\ B {\bf 485}, 85 (1997)
[arXiv:hep-th/9608024].

\bibitem{Aharony:2000ti}
O.~Aharony, S.~S.~Gubser, J.~Maldacena, H.~Ooguri and Y.~Oz,
Phys.\ Rept.\  {\bf 323}, 183 (2000)
[arXiv:hep-th/9905111].


\bibitem{Cardy:1989ir}
J.~L.~Cardy,
Nucl.\ Phys.\ B {\bf 324}, 581 (1989).

\bibitem{Lewellen:1992tb}
D.~C.~Lewellen,
Nucl.\ Phys.\ B {\bf 372}, 654 (1992).

\bibitem{Cardy:1991tv}
J.~L.~Cardy and D.~C.~Lewellen,
Phys.\ Lett.\ B {\bf 259}, 274 (1991).

\bibitem{Pradisi:1996yd}
G.~Pradisi, A.~Sagnotti and Y.~S.~Stanev,
Phys.\ Lett.\ B {\bf 381}, 97 (1996)
[hep-th/9603097].

\bibitem{Fuchs:1998fu}
J.~Fuchs and C.~Schweigert,
Nucl.\ Phys.\ B {\bf 530}, 99 (1998)
[hep-th/9712257].

\bibitem{Bachas:2000ik}
C.~Bachas, M.~R.~Douglas and C.~Schweigert,
JHEP {\bf 0005}, 048 (2000)
[hep-th/0003037].

\bibitem{Maldacena:2001ky}
J.~Maldacena, G.~W.~Moore and N.~Seiberg,
JHEP {\bf 0107}, 046 (2001)
[hep-th/0105038].


\bibitem{Bachas:2001bt}
C.~Bachas,
hep-th/0106234. and references therein

\bibitem{Dudas:2001wd}
E.~Dudas, J.~Mourad and A.~Sagnotti,
hep-th/0107081.


\bibitem{Melvin}
M. Melvin,
Phys. Lett.
8 (1964) 65.

\bibitem{Gibbons:1988ps}
G.~W.~Gibbons and K.~Maeda,
Nucl.\ Phys.\ B {\bf 298}, 741 (1988).
\bibitem{Dowker:1995gb}
F.~Dowker, J.~P.~Gauntlett, G.~W.~Gibbons and G.~T.~Horowitz,
Phys.\ Rev.\ D {\bf 52}, 6929 (1995)
[hep-th/9507143].


\bibitem{Dowker:1994up}
F.~Dowker, J.~P.~Gauntlett, S.~B.~Giddings and G.~T.~Horowitz,
Phys.\ Rev.\ D {\bf 50}, 2662 (1994)
[hep-th/9312172].

\bibitem{Dowker:1994bt}
F.~Dowker, J.~P.~Gauntlett, D.~A.~Kastor and J.~Traschen,
Phys.\ Rev.\ D {\bf 49}, 2909 (1994)
[hep-th/9309075].

\bibitem{Dowker:1996sg}
F.~Dowker, J.~P.~Gauntlett, G.~W.~Gibbons and G.~T.~Horowitz,
Phys.\ Rev.\ D {\bf 53}, 7115 (1996)
[hep-th/9512154].

\bibitem{Russo:1995tj}
J.~G.~Russo and A.~A.~Tseytlin,
Nucl.\ Phys.\ B {\bf 449}, 91 (1995)
[hep-th/9502038].

\bibitem{Russo:1996ik}
J.~G.~Russo and A.~A.~Tseytlin,
Nucl.\ Phys.\ B {\bf 461}, 131 (1996)
[hep-th/9508068].

\bibitem{Costa:2001nw}
M.~S.~Costa and M.~Gutperle,
JHEP {\bf 0103}, 027 (2001)
[hep-th/0012072].

\bibitem{Gutperle:2001mb}
M.~Gutperle and A.~Strominger,
JHEP {\bf 0106}, 035 (2001)
[hep-th/0104136].
\bibitem{Costa:2001if}
M.~S.~Costa, C.~A.~Herdeiro and L.~Cornalba,
hep-th/0105023.

\bibitem{Emparan:2001rp}
R.~Emparan,
Nucl.\ Phys.\ B {\bf 610}, 169 (2001)
[hep-th/0105062].

\bibitem{Saffin:2001jg}
P.~M.~Saffin,
hep-th/0105220.

\bibitem{Suyama:2001bn}
T.~Suyama,
JHEP {\bf 0108}, 037 (2001)
[hep-th/0106079].
T.~Suyama,
hep-th/0107116.
T.~Suyama,
hep-th/0110077.

\bibitem{Adams:2001sv}
A.~Adams, J.~Polchinski and E.~Silverstein,
hep-th/0108075.


\bibitem{Uranga:2001dx}
A.~M.~Uranga,
hep-th/0108196.

\bibitem{Russo:2001na}
J.~G.~Russo and A.~A.~Tseytlin,
hep-th/0110107.

\bibitem{Takayanagi:2001jj}
T.~Takayanagi and T.~Uesugi,
hep-th/0110099.


\bibitem{boundary}
C.~G.~Callan, C.~Lovelace, C.~R.~Nappi and S.~A.~Yost,
Nucl.\ Phys.\ B {\bf 293}, 83 (1987);
J.~Polchinski and Y.~Cai,
Nucl.\ Phys.\ B {\bf 296}, 91 (1988);
N.~Ishibashi,
Mod.\ Phys.\ Lett.\ A {\bf 4}, 251 (1989).


\bibitem{ss}
J.~Scherk and J.~H.~Schwarz,
Nucl.\ Phys.\ B {\bf 153} (1979) 61.

\bibitem{ssclosed}
R.~Rohm,
Nucl.\ Phys.\ B {\bf 237} (1984) 553;
S.~Ferrara, C.~Kounnas and M.~Porrati,
Phys.\ Lett.\ B {\bf 181} (1986) 263,
Nucl.\ Phys.\ B {\bf 304} (1988) 500,
Phys.\ Lett.\ B {\bf 206} (1988) 25;
C.~Kounnas and M.~Porrati,
Nucl.\ Phys.\ B {\bf 310} (1988) 355;
S.~Ferrara, C.~Kounnas, M.~Porrati and F.~Zwirner,
Nucl.\ Phys.\ B {\bf 318} (1989) 75;
C.~Kounnas and B.~Rostand,
Nucl.\ Phys.\ B {\bf 341} (1990) 641;
I.~Antoniadis and C.~Kounnas,
Phys.\ Lett.\ B {\bf 261} (1991) 369;
E.~Kiritsis and C.~Kounnas,
Nucl.\ Phys.\ B {\bf 503} (1997) 117
[hep-th/9703059].

\bibitem{ssopen}
I.~Antoniadis, E.~Dudas and A.~Sagnotti,
Nucl.\ Phys.\ B {\bf 544} (1999) 469
[hep-th/9807011];
I.~Antoniadis, G.~D'Appollonio, E.~Dudas and A.~Sagnotti,
Nucl.\ Phys.\ B {\bf 553} (1999) 133
[hep-th/9812118];
Nucl.\ Phys.\ B {\bf 565} (2000) 123
[hep-th/9907184];
R.~Blumenhagen and L.~Gorlich,
Nucl.\ Phys.\ B {\bf 551} (1999) 601
[hep-th/9812158];
C.~Angelantonj, I.~Antoniadis and K.~Forger,
Nucl.\ Phys.\ B {\bf 555} (1999) 116
[hep-th/9904092].



\end{thebibliography}
\end{document}